\documentclass[runningheads,screen]{llncs}
\usepackage[LNCS,
   key=2021-Werner-ICSOC-CoDesign,
   year=2021,
   publication={S. Werner and S. Tai, Proceedings of the 19th International Conference on Service-Oriented Computing, ICSCO'21.}
   Preprint Version,
   nourl,
   nobib
 ]{authorarchive}

\input{src/packages}
\definecolor{darkblue}{rgb}{0.122, 0.435, 0.698}
\definecolor{grey}{gray}{0.30}
\definecolor{lightblue}{rgb}{0, 0.8, 0.99}

\definecolor{sky1a}{HTML}{cfe0e8}
\definecolor{sky1b}{HTML}{b7d7e8}
\definecolor{sky1c}{HTML}{87bdd8}
\definecolor{sky1d}{HTML}{daebe8}

\definecolor{sky2a}{HTML}{bccad6}
\definecolor{sky2b}{HTML}{8d9db6}
\definecolor{sky2c}{HTML}{667292}
\definecolor{sky2d}{HTML}{f1e3dd}

\definecolor{living_coral}{HTML}{FF6F61}
\definecolor{spearmint}{HTML}{64BFA4}

\definecolor{teal}{HTML}{00A08F}
\definecolor{marine}{HTML}{124653}
\definecolor{yellow}{HTML}{FEE074}
\definecolor{orange}{HTML}{FF9469}
\definecolor{pink}{HTML}{FE8d8F}

\makeatletter
\@ifclasswith{llncs}{review}{
\renewcommand{\author}[1]{}
\renewcommand{\institute}[1]{}
}{}

\@ifclasswith{llncs}{review}{
    \newcommand{\hcite}[1]{[removed]}
}{
    \newcommand{\hcite}[1]{\cite{#1}}
}

\@ifclasswith{llncs}{review}{
    \newcommand{\hide}[1]{(removed)}
}{
    \newcommand{\hide}[1]{#1}
}
\makeatother

\begin{document}
\title{Application-Platform Co-Design for\\  Serverless Data Processing}
\titlerunning{Application-Platform Co-Design for Serverless Data Processing}

\author{Sebastian Werner\inst{1} \and
Stefan Tai\inst{1}}
\authorrunning{\hide{S. Werner \and S. Tai}}

\institute{Information Systems Engineering - Technische Universit\"at Berlin, Berlin, Germany\\
\email{\{sw,st\}@ise.tu-berlin.de}}
\maketitle              %
\begin{abstract}
“Application-platform co-design” refers to the phenomenon of new platforms being created in response to changing application needs, followed by application design and development changing due to the emergence (and the specifics, limitations) of the new platforms, therefore creating, again, new application and platform requirements. This continuous process of application and platform (re-)design describes an engineering and management responsibility to constantly evaluate any given platform for application fit and platform-specific application design, and to consider a new or evolutionary platform development project due to evolving and changing application needs. 

In this paper, we study this phenomenon in the context of serverless computing and (big) data processing needs, and thus, for application-platform co-design for serverless data processing (SDP). We present an analysis of the state-of-the-art of function-as-a-service (FaaS) platforms, which reveals several configuration, deployment, execution, and measurement differences between popular platforms happening at-speed. These differences indicate already ongoing platform (re-)design processes resulting in more specialized serverless platforms and new, platform-specific challenges for application design.
We discuss data processing needs of applications using the serverless model and present common initial (and undesirable) workaround solutions on the application level, giving additional argument to the creation of new SDP platforms. We present critical SDP requirements and possible new platform augmentations, but identify the need for engineering methods and tooling to better guide application-platform co-design. 
We argue to pay appropriate attention to the phenomenon of continuous application-platform co-design to better anticipate and to control future platform and application developments.
\end{abstract}

\section{Introduction}

Traditionally, new software platforms were created in response to new application demands, such as specific elasticity or big data processing requirements. Once a new platform is in place, application design and development on top of the platform has to take the platform features, specifics, and constraints into account. Often, the impact of a new platform on application design and development is significant. And in turn, new application requirements are created as a result, which, again, may suggest the development of a new (variant of the) software platform. 

Notably, the advent of NoSQL database systems serves as an example for this phenomenon of application-platform co-design. Originally initiated by data processing and concurrency needs of large enterprises, system designs like, Google's GFS~\cite{gfs} and Amazon's DynamoDB~\cite{dynamo}   fueled an explosion of numerous (ca. over 250) new, NoSQL data storage platforms over the last decades. Today, a developer can choose between a magnitude of managed and self-managed data storage systems that can meet almost every niche application requirement. But each platform, however, may provide different data consistency guarantees, shifting data synchronization or conflict resolution, for example, from the platform to the application as a new application responsibility.

Analogously, the way we run applications on cloud platforms has been evolving significantly and at-speed, too. Web services can be deployed on elasticity managed VMs, with sophisticated container orchestration platforms such as Kubernetes, or using tiny micro-VMs in a serverless setting. Modern cloud platforms, thus, already support a plethora of ways a developer can deploy, scale, and run web-serving applications.

The same application-platform co-design phenomenon can be observed, too, within the field of serverless computing. With serverless computing, the basic idea is to free application developers from responsibilities related to elasticity, deployment, and monitoring, that is, from almost any operational task. Current serverless platforms, specifically Function-as-a-Service (FaaS) offerings, have rapidly changed and improved since their early introduction in 2014. The initial one-size-fits-all model suggested with serverless computing has already, almost in the background, started to shift, and several variants of serverless platforms serving different application needs than just simple web-serving tasks have emerged~\cite{feshen,SMILE}. 

Specifically, distributed data processing~\cite{2019-Carver-PDSW-Wukong,2019-Fouladi-USENIX-gg,2020-Mueller-SIGMOD-Lambada} shows to benefit from the serverless computing model and its extreme scalability, low operational overhead, work-based billing model, and overall simplicity. Moreover, classical data processing frameworks, e.g., Apache Spark, Hive, and Apache Flink, require data analysts to deploy, configure, and operate clusters of servers and thus require developer responsibility for operational tasks that can impose considerable and potential disastrous entry barriers~\cite{2018-Werner-BigData-MatrixMultiplication} to anyone that needs to analyze data. Serverless data processing frameworks, such as Lithops~\cite{2021-Sampe-Software-Lithops} and Pywren~\cite{2017-Jonas-SoCC-PyWren} aim to reduce such entry barriers by providing data processing APIs with similar abstractions to classical frameworks  without upfront cluster management needs.

Be it NoSQL stores, cloud platforms, FaaS offerings, or data processing solutions, the continuous cycle of application-platform co-design has led and is still leading to an abundance of platforms, some of which differ only in details, and some of which differ significantly. This introduces the continuous need to question the application fit of any given platform, to design applications in a platform-specific manner, or, to develop a new general-purpose or application-specific platform (variant). As a consequence, software engineering requires increasing attention to be paid to the diverse phenomena of application-platform co-design.

In this paper, we study and discuss application-platform co-design for serverless data processing. Based on an analysis of the current state of serverless platforms, we highlight areas where current platforms are already differentiating themselves from each other. Further, we discuss data processing needs of applications using the serverless computing model and present common initial, but undesirable workaround solutions on the application level, giving additional argument to the creation of new serverless data processing (SDP) platforms. We present critical SDP requirements and possible new platform augmentations, but identify the need for engineering methods and tooling to better guide application-platform co-design. We argue to pay appropriate attention to the phenomenon of continuous application-platform co-design.

\section{Serverless Computing Platforms}\label{sec:platforms}
Let us first take a closer look at the current state of serverless computing platforms. In this section, we specifically compare the popular FaaS offerings of the four major cloud providers Amazon, Google, Microsoft and IBM, and highlight both similarities and differences.

Cloud-based FaaS offerings, the most widely adopted form of serverless computing, ask developers only to define applications through arbitrary function code and triggering event definitions.
The cloud provider is responsible for deploying, running, and scaling these functions in response to arriving events.
For all cloud providers, developers can select from a set of predefined runtime environments and only manage few additional configurations, such as setting memory limits, maximum concurrency and environment variables.
Thus, all current FaaS offerings enable almost operations free delivery of stateless serverless applications. 
However, current offerings still lack support for state management, hardware acceleration and suitable programming abstractions~\cite{2020-Kuhlenkamp-IC2E-ifs_and_buts,2021-schleier-acm-next-phase} to support any cloud-based application, although platform vendors already started to differentiate themselves by addressing these and other open serverless challenges~\cite{2019-Jonas-arxiv-BerkeleyViewOnServerlessComputing}.

At first sight, from a developer perspective, all platforms provide a similar programming interface and execution model.
Thus, in theory, the choice of a specific serverless computing platform should not significantly affect the application design.
However, taking a closer look, the available configuration space, runtime isolation, platform limitations and auxiliary services can differ substantially between the different platforms, and thus, careful developer consideration is a must.

Further, larger applications built as serverless systems do not consist of a single function but a composition of functions and other services. The available platform services for function composition and orchestration differ significantly.

\subsection{Platform Comparison}
Table~\ref{tab:platforms} provides a comparative overview of the serverless computing platforms from Amazon (AWS), Google (GCF), IBM (ICF) and Microsoft (ACF). We compare these platforms along four general categories: First, configuration options -- all exposed ``tuning knobs'' a developer can control; 
second, deployment options -- e.g., available runtimes and deployment environments;
and third, execution criteria -- important criteria for function execution and existing limits.
Finally, we also provide some basic metrics and measurements that indicate platform qualities such as performance or elasticity. 

\begin{table}[t]
\caption{Configuration, deployment, execution, and measurement differences in FaaS}
\label{tab:platforms}
\resizebox{\textwidth}{!}{%
\begin{tabular}{|
>{\columncolor[HTML]{EFEFEF}}l |
>{\columncolor[HTML]{EFEFEF}}r |r|r|r|r|l|}
\hline
\multicolumn{2}{|l|}{\cellcolor[HTML]{EFEFEF}} & \multicolumn{1}{l|}{\cellcolor[HTML]{EFEFEF}AWS} & \multicolumn{1}{l|}{\cellcolor[HTML]{EFEFEF}GCF} & \multicolumn{1}{l|}{\cellcolor[HTML]{EFEFEF}ICF} & \multicolumn{1}{l|}{\cellcolor[HTML]{EFEFEF}ACF} & \cellcolor[HTML]{EFEFEF}Source \\ \hline
\cellcolor[HTML]{EFEFEF} & min. Mem. {[}MB{]} & 128 & 128 & 128 & N.A. & Docs \\ \cline{2-7} 
\cellcolor[HTML]{EFEFEF} & max. Mem.  {[}MB{]} & 10240 & 8192 & 2048 & 14336 & Docs \\ \cline{2-7} 
\cellcolor[HTML]{EFEFEF} & Memory Space {[}\#{]} & 10112 & 7 & 1920 & 14336 & Docs \\ \cline{2-7} 
\cellcolor[HTML]{EFEFEF} & Timeout {[}s{]} & 900 & 540 & 60 & 600 & Docs \\ \cline{2-7} 
\cellcolor[HTML]{EFEFEF} & vCPU Cores {[}\#{]} & 6 & 1 & ? & 4 & Docs \\ \cline{2-7} 
\cellcolor[HTML]{EFEFEF} & CPU {[}GHz{]} & 2.5 & 4.8 & ? & 2.4 & Docs \\ \cline{2-7} 
\multirow{-8}{*}{\cellcolor[HTML]{EFEFEF}\rotatebox[origin=c]{90}{Configuration}} & max Concurrency {[}\#{]} & 1000+ & 1000 & 1000+ & VM*100 & Docs \\ \hline
\cellcolor[HTML]{EFEFEF} & Trigger {[}\#{]} & 8 & 6 & 3+ & 4+ & Docs \\ \cline{2-7} 
\cellcolor[HTML]{EFEFEF} & Supported Runtimes {[}\#{]} & 15+ & 13 & 8+ & 7 & Docs \\ \cline{2-7} 
\cellcolor[HTML]{EFEFEF} & Dependency Management & Layers & Files & Docker/Files & Files & Docs \\ \cline{2-7} 
\cellcolor[HTML]{EFEFEF} & max. Size {[}MB{]} & 250 & 500 & 48 & -1 & Docs \\ \cline{2-7} 
\multirow{-5}{*}{\cellcolor[HTML]{EFEFEF}\rotatebox[origin=c]{90}{Deployment}} & Host Controllable & No & No & No & Yes & Docs \\ \hline
\cellcolor[HTML]{EFEFEF} & Isolation & firecracker & gVisor & VM+runc & VM & Docs \\ \cline{2-7} 
\cellcolor[HTML]{EFEFEF} & Event Scheduling & push-based & unknown & push-based & pull-based & \cite{2021-Barcelona-FGCS-Parallel} \\ \cline{2-7} 
\cellcolor[HTML]{EFEFEF} & Local Storage {[}MB{]} & 512 & 0 & 0 & 143360 & Docs \\ \cline{2-7} 
\cellcolor[HTML]{EFEFEF} & Network Storage {[}Y/N{]} & Yes & No & No & Yes & Docs \\ \cline{2-7} 
\cellcolor[HTML]{EFEFEF} & Private Networking {[}Y/N{]} & Yes & Yes & No & Yes & Docs \\ \cline{2-7} 
\cellcolor[HTML]{EFEFEF} & Function Networking {[}Y/N{]} & Unsupported & Unsupported & Unsupported & Yes & Docs \\  \cline{2-7} 
\cellcolor[HTML]{EFEFEF} & Tracing {[}Y/N{]} & Yes & Yes & No & No & \cite{2021-Borges-IC2E-Debuging} \\ \cline{2-7} 
\cellcolor[HTML]{EFEFEF} & Function Metrics {[}Y/N{]} & Yes & Yes & Yes & Yes & \cite{2021-Borges-IC2E-Debuging} \\ \cline{2-7} 
\cellcolor[HTML]{EFEFEF} & Cloud Logs {[}Y/N{]} & Yes & Yes & Yes & Yes & \cite{2021-Borges-IC2E-Debuging} \\ \cline{2-7} 
\cellcolor[HTML]{EFEFEF} & Billing Interval {[}time{]} & 1ms & 100ms & 100ms & 100ms-1hour & Docs \\ \cline{2-7} 
\cellcolor[HTML]{EFEFEF} & Threads {[}\#{]} & 1024 & unknown & 1024 &  varies  & Docs \\ \cline{2-7} 
\cellcolor[HTML]{EFEFEF} & Connections {[}\#{]} & 1024 & unknown & 1024 & 600 & Docs \\ \cline{2-7} 
\cellcolor[HTML]{EFEFEF} & Payload Size {[}MB{]} & 6 & 10 &  & 100 & Docs \\ \cline{2-7} 
\multirow{-14}{*}{\cellcolor[HTML]{EFEFEF}\rotatebox[origin=c]{90}{Execution}} & Rate Limit & 10x1000 /s & 100MB/s & 84/s & unmanaged & Docs \\ \hline
\cellcolor[HTML]{EFEFEF} & Configiruation Chages {[}ms{]} & 996 & 36630 & 22 & 521100 & \cite{2019-Kuhlenkamp-Ucc-Opstasks}\\ \cline{2-7} 
\cellcolor[HTML]{EFEFEF} & Cold Start Variance {[}ms{]} & 9 & 4900 & 10528 & 83691 & \cite{2020-Kuhlenkamp-ACR-All_But_One} \\ \cline{2-7} 
\multirow{-3}{*}{\cellcolor[HTML]{EFEFEF}\rotatebox[origin=c]{90}{Meas.}} & Cold Default Throughput {[}trps{]} & 120 & 120 & 120 & 5 & \cite{2020-Kuhlenkamp-ACR-All_But_One} \\  \hline
\end{tabular}%
}
\end{table}

\paragraph{Configuration} reveals two principle models: AWS, GCF, and ICF expose developers to a singular, highly sensitive performance-related sizing parameter. On the other end, AWS offers over 10.000 unique settings to control performance. GCF, in contrast, presents only seven options. This singular parameter affects multiple resource sizes simultaneously, e.g., memory, network bandwidth, available threads. 
Hiding many complex resource configurations behind a singular value leads to the need for sizing tools.
With Azure, however, the ability to select from different VM offerings as a back-end for serverless workloads exists and so, the sizing problem is different.

\paragraph{Deployment} options are similar for all platforms under comparison. While the number of selectable runtimes differs, the most common programming languages are supported by all platforms.
A major difference, however, relates to dependency management. 
The limited allowance for deployment package sizes (between 50-500MB) and the management of dependency versions has led some platform provides to offer more advanced features for dependency management. 
Among them, AWS allows developers to build shareable layers that multiple functions can reuse. IBM's OpenWhisk opened the runtime API to enable developers to define complete docker images with all dependencies built-in to address this issue.

\paragraph{Execution} in serverless computing platforms is based on three main factors: Function isolation, assignment of invocations (execution guarantees), and invocation triggering. 

For isolation, AWS uses firecracker\cite{2020-Agache-NSDI-Firecracker}, a KVM based micro-VM. Thus, each function is strongly isolated while removing comparably long startup times of classical VMs.
Google uses gVisor, a form of OS-level isolation that shares common roots with AWS firecracker but is also used for other Google services and thus is less specialized. 
ICF and Azure use a VM per user to isolate functions. Thus, functions might interfere with the execution of other functions of the same user while not interfering with functions of other users. Here, the scaling of functions depends on the time it takes to launch new VMs per customer. 

Besides isolation, the assignment of events to functions is different between these platforms. For AWS, GCF and ICF, we see a pull-based approach: free hosts will pull available events. 
Azure, on the other hand, uses a push-based approach, which can impact elasticity.

Lastly, all platforms offer means to trigger functions synchronously and asynchronously. However, the number of available options to trigger functions can differ. For instance, AWS provides triggers for most database services. At the same time, other platforms such as ACF or ICF give developers only a few endpoints to trigger functions synchronously or asynchronously.

\subsection{Vendor Directions}
The comparison shows that the current landscape of serverless platforms shares a common programming and operations model, while at the same time, revealing notable differences with respect to limitations and configurable resources between platforms.

Some recent platform (re-)design efforts taken by cloud providers further include introductions of additional platform services and features to overcome identified shortcomings. For example, Microsoft recently introduced durable functions, a programming model to store function states after execution. 
Similarly, Amazon recently added the Elastic File System (EFS) for Lambda, thus enabling functions to persist data across multiple executions, multiple function-deployments and between parallel invocations.

Vendors are constantly differentiating their offerings and as a consequence, the initial common programming model shared between multiple platforms diverges into diverse, different models, making it nearly impossible to switch platforms later on.
Moreover, the larger serverless research and practitioners' community has started to propose novel changes for FaaS platforms as well, addressing some of the most commonly identified serverless shortcomings~\cite{2019-Hellerstein-CIDR-ServerlessComputingSteps}, again resulting in diverse platform developments.

\section{Serverless Data Processing}\label{sec:sdp}
Let us now look into modern applications' data processing needs and how these translate into serverless data processing (SDP) requirements.

\subsection{System Requirements}
We conducted a series of experiments related to serverless computing and (big) data processing, initially presented in 2018~\hcite{2018-Werner-BigData-MatrixMultiplication} and continued with~\hcite{2018-Kuhlenkamp-WoSC-Survey,2019-Kuhlenkamp-Ucc-Opstasks,2020-Kuhlenkamp-ACR-All_But_One} and~\hcite{2020-Werner-WoSC-SBDPs}. From these lessons learned, we define the following serverless data processing system (platform and application) requirements:

\begin{enumerate}%
    \item \textbf{Scaleable}: A serverless data processing system should use the scalability potential of a serverless platform and adapt the resource demands of each computation to the task. Further, the system should have comparable performance characteristics as conventional data processing solutions (such as an Apache Spark Cluster) of similar cost and size.
    
    \item \textbf{Fully-Serverless}: The serverless data processing system should be fully serverless, that is, the serverless data processing system should be able to scale down to zero if no resources are needed. Thus, the system should not incur costs or management tasks if idle (an exception can be made for storing input data).
    Further, the analyst should not know the inner workings of the used services, such as avoiding cold-starts or selecting the optimal size of AWS S3 files for Lambda. 
    
    \item \textbf{Self-Contained}: The system should be self-contained. Specifically, the system should handle deployment, re-execution of faulty invocations or re-configuration of wrongly sized execution environments. 
    
    \item \textbf{Tuneable}: The system should allow developers to define high-level objectives for each computation, such as low cost or fast computation time. The system should drive all configurations and executions based on these high-level tunables.
    
    \item \textbf{Integratable}: Modern data processing applications need to combine multiple tools, programs and algorithms for pre/post-processing to appropriately integrate with all relevant business processes. Thus, the system should allow for arbitrary, yet performance-aware pre/post-processing integration.
    
\end{enumerate}

Serverless data processing platforms should be as versatile as conventional data processing solutions such as Apache Spark or Apache Flink.
However, not all use-cases will benefit equally from the properties of serverless data processing~\cite{2019-Hellerstein-CIDR-ServerlessComputingSteps}. 
We observe that SDP is most useful for ad-hoc analytics~\hcite{2020-Werner-WoSC-SBDPs}, tasks such as data cleaning, data inspections, as well as IoT scenarios such as predictive maintenance or troubleshooting. 
Similarly, exploratory data analytics relevant in pre-processing of machine learning~\cite{2020-Schifferer-RSC-GPU} can benefit well from the ad-hoc processing capabilities of SDP.
Further, tasks that only require infrequent processing, such as indexing for data lakes, also benefit from the fast deployment and re-deployment of processing resources in the serverless model.

\subsection{Common Application Workarounds}\label{sec:workarounds}
Multiple SDP frameworks have emerged in the last four years~\cite{2019-Carver-PDSW-Wukong,2017-Jonas-SoCC-PyWren,2020-Mueller-SIGMOD-Lambada,2021-Sampe-Software-Lithops}.
Naturally, the complexity of available programming interfaces has increased and different options exist to address the serverless data processing requirements identified above.

\noindent
The most common trend, however, still present in all SDP frameworks, is the use of workarounds to overcome known platform limitations.

\paragraph{\textbf{Serverless job orchestration}} involves the generation of invocations for each task in a processing job, waiting on the completion of these invocations and the collection or redistribution of task results. 
Each of these steps can be addressed in different ways. 
A driver can generate events asynchronously (the most common approach), thus, only submitting tasks to the serverless platform. 
In that case, the driver now has to query the platform repeatedly to observe each task. 
This design forces an extensive network and request overhead to enable drivers to observe functions in real-time.

Alternatively, each invocation can be performed synchronously, removing the need to constantly poll for results but, in turn, limiting the maximum number of concurrent invocations a single driver can manage. 
Most platforms require that each synchronous invocation contains a single event and thus requires a driver to open as many connections as functions should run in parallel. With this strategy, it is virtually impossible to reach the scalability potential of state of the art serverless platforms. 

A third option is to use a platform-specific orchestration mechanism, such as a workflow engine, for example, AWS Step-Functions. 
However, current platform-specific orchestration mechanisms are all geared for orchestrating a flow of events through a tree of different functions rather than facilitating a highly parallel execution of few functions. 
On top of that, each mechanism increases management and configuration overhead and makes migration to other platforms far more work-intensive.

A fourth strategy that we see is to spawn functions without specific instructions. Instead, each function connects to an external service to pull tasks from a shared task queue~\cite{2019-Carver-PDSW-Wukong}.
This approach removes the need to observe the completion of functions through the serverless platform and can use a lightweight mechanisms to launch many functions in parallel. However, at the cost of introducing new external dependencies that are difficult to maintain, to scale and that typically are not serverless, fundamental requirements of serverless data processing may be broken.

\paragraph{\textbf{Serverless state management}} in serverless data processing is divided into two major sub-problems, intermediate storage and data access.
Essentially, both intermediate data storage, data ingestion and saving results involve an external state management system.
Here, frameworks commonly use an object store such as Amazon's S3 or a managed message queue system.
However, these systems introduce latency and network overhead for each computation. 
As an added complication, each function has to manage the connection to the storage regardless of the selected back end, thus, introducing added overhead per function and many more sources for errors to occur.
It is further unclear if the selected storage-backends are well suited to transport the type of ephemeral data efficiently.

\paragraph{\textbf{Serverless uniformity}} also creates a challenge for framework designers. 
In most cases, a framework will deploy one function per task or sometimes even a single function for all tasks in a processing job.
Thus, the sizing of that function must always fit the largest part of a task to ensure that a computation does not run out of memory or takes too long.
Consequently, serverless processing systems either struggle with processing skew or otherwise heterogeneous data or waste a significant amount of resources. 
The fact that platforms do not allow applications to implement custom failure recovery mechanisms, such as temporary increasing resource limits, to address these issues means that application developers need to find other solutions. 

Additionally, we observe that the cold-start of functions is impacted by both memory size selection and deployment package size~\cite{2018-Manner-WoSC-Benchmarking_Cold_Starts}.
Thus, the design of current SDP frameworks must take both runtime size and sizing into account to address cold-start issues. 
Therefore, it should come as no surprise that most of these frameworks target AWS Lambda, as it is the most flexible platform in terms of runtime environments, deployment sizes, and memory sizes. 
However, it remains to be seen if platform improvements can be equally or even better provided for Azure, Google or IBM SDP platforms.

\paragraph{\textbf{Serverless support eco-system}} describes the problem of selecting appropriate services to augment missing features in the serverless compute platform. 
Most frameworks rely on one or more additional cloud-based infrastructure services to fully support each processing step.
Thus, selecting a suitable service can often impact the overall performance, manageability and cost of a framework.
These auxiliary services often differ significantly between vendors, making the portability of these frameworks problematic as well. 
Moreover, are these auxiliary services are rarely designed for serverless workloads and serverless data processing workloads. 
In particular, the usage for orchestration or data transfer is often inefficiently supported or could easily break if vendors decide to change service properties without serverless workloads in mind.
Consequently, the design of serverless data processing systems is strongly dependent on the selected cloud platform and the composition of available auxiliary services and serverless computing resources.

\subsection{Next Steps}
Current SDP applications, unsurprisingly, already utilize existing FaaS platforms and SDP frameworks quite well. 
However, as discussed above, there are many specifics and platform and programming model limitations that quickly lead to potentially significant design inefficiencies and platform lock-in. 
Applications need to adapt to platform evolution and welcome desirable innovations, such as higher-level programming abstractions. While early SDP frameworks only supported bare-bone map-reduce, the more recent frameworks start to support higher-level APIs and query languages. 
Nevertheless, the prominent presence of many workarounds as described above, and the use of auxiliary services that were never intended to serve as a backbone to highly parallel computations, creates a significant risk regarding the usage of current SDP frameworks.

\section{Towards Guided Co-Design}\label{sec:apcd}
We expect serverless platforms and current serverless data processing applications to continue to evolve to fully support all serverless data processing requirements. To this end, we envision current limitations and workarounds to be replaced by solutions that require new platform augmentations. At the same time, we see the need for new engineering methods and tooling to better guide platform and application re-design and evolution.

\subsection{New Platform Augmentations}
We can identify function orchestration, intermediate data transfer, and straggling executions as the most pressing issues in the SDP context requiring new platform augmentations.
In the following, we revisit the undesirable workarounds presented in Section~\ref{sec:workarounds} and discuss how platform augmentations, or the selection of new platform features, can remove these issues while remaining true to the serverless data processing model.

Our discussion and recommendations are based on own prior and other related work, including both exploratory FaaS studies~\cite{2018-Kuhlenkamp-WoSC-Survey,2018-Leitner-JoSS-FaasInPractice,2020-Werner-XP-Diminuendo}, benchmarks~\cite{2019-Kuhlenkamp-Ucc-Opstasks,2018-Manner-WoSC-Benchmarking_Cold_Starts} and technical platform papers~\cite{2020-Agache-NSDI-Firecracker} as well as emerging open-source developments~\cite{feshen}, SDP prototype developments~\cite{2018-Werner-BigData-MatrixMultiplication} and exploratory SDP studies~\cite{2020-Werner-WoSC-SBDPs}. 

For \textbf{serverless job orchestration}, the different approaches discussed all can introduce undesirable inefficiencies. Each of the presented workarounds thus introduces a possible adaption cause. 

Based on benchmarks performed in previous work~\cite{2020-Kuhlenkamp-ACR-All_But_One}, it appears that AWS is the most suitable platform for using synchronous executions. Alternatively, we can augment existing platforms to address the issue of spawning and observing multiple function invocations simultaneously.
For example, platforms can introduce new means to batch invocations with a callback on completion to allow frameworks to spawn thousands of functions without the need to manage each invocation individually.
Thus, this immediately removes the need to create complicated management structures around existing platform APIs from a developer perspective. This would also allow for more predictive scheduling and reduce overly aggressive polling of APIs for these types of use-cases from a platform perspective.

For \textbf{serverless state management} several proposals to address the intermediate storage problem are already emerging.
Klimov et al. \cite{2018-Klimovic-ATC-Ephemeral_Storage}, for example, propose flash-based storage that can be used by serverless analytics in place of the currently used object storage for intermediate data.
However, platforms could aid function developers by offering an intermediate storage layer on each worker to address intermediate storage needs on a platform level. These could hold data for a short time, thus allowing functions to batch read, write to external data sources, or even reuse data for intermediate computations.
Further, platforms could address the problem of redundant connection to the selected storage back-end by integrating connection pooling on the worker level.

\textbf{Serverless uniformity} can in part be addressed on the application level by chaining the deployment strategy of current frameworks. Instead of deploying a function with only one configuration, frameworks could deploy functions in multiple sizes and switch the invocations to larger deployments in case of skewed data. 
However, not all platforms allow flexible sizing of deployment packages, and thus, developers risk oversizing and overpaying with this strategy.
Here, platforms can offer more flexible sizing options, integrate sizing aid at runtime or enable other mechanisms to adjust deployments in case of errors.

Furthermore, the programming interface of functions could be extended to include other life-cycle related events such as function termination and function-creating to allow frameworks to group some common tasks on the start and end of a function life-cycle instead of every single execution, thus reducing the risk of timeouts during IO operations. Also, we foresee new serverless platforms that are breaking even more with the initial one-size-fits-all model of serverless computing to address specific application requirements, such as the support of computation accelerators~\cite{2021-Pedro-arxiv-Predictions}, edge-computing infrastructure and optimized systems for parallel computing.

\subsection{Understanding Co-Design}
As discussed in general in the introduction and as exemplified for serverless data processing, software platforms will continue to evolve or be newly created in response to changing application demands. These platforms continue to push the envelope of what applications can do and thus again present new demands that motivate platform changes and, ultimately, again lead to new platforms.

This phenomenon of application platform co-design takes place both consciously and unconsciously between platform and application developers. Understanding this phenomenon better enables application developers to anticipate platform changes as well as new platforms, and thus allows for better management of coming changes.
Similarly, application developers can take control of the application-platform co-design cycle and influence new platform developments directly.

\begin{figure}
    \centering
    \includegraphics[width=\columnwidth]{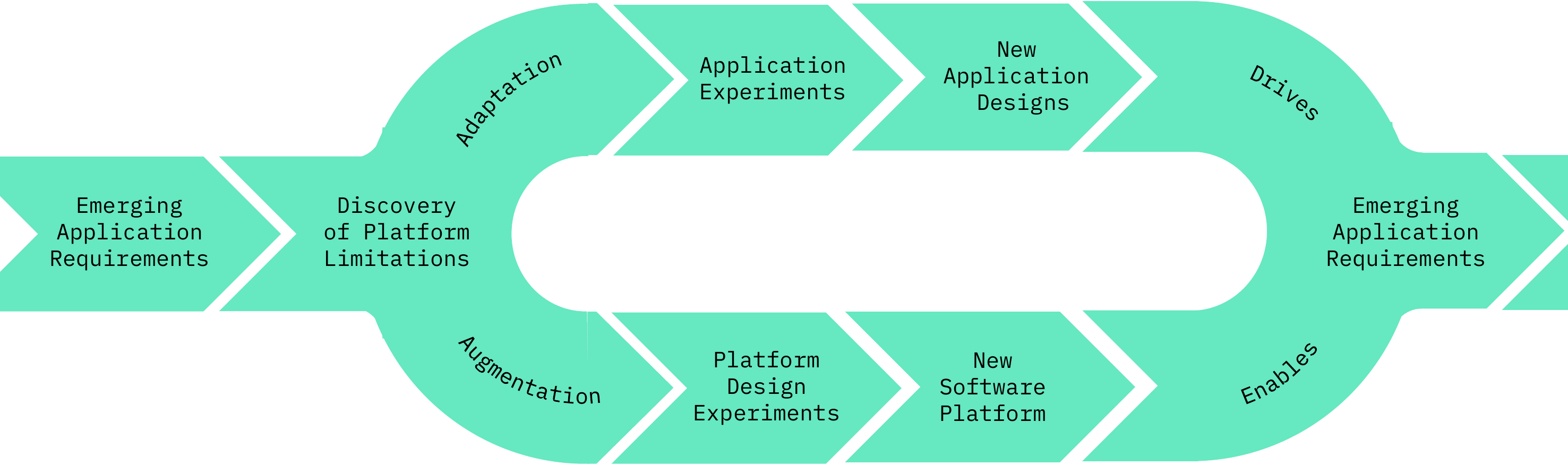}
    \caption{Conceptual view of application platform co-design cycle}
    \label{fig:method}
\end{figure}

Figure~\ref{fig:method} illustrates the continuous nature of application platform co-design.
Newly emerging application demands and requirements drive the discovery of software platform limitations, for example, native state management in serverless computing. 
Once a limitation is known, application developers start to use workarounds, as described in section~\ref{sec:workarounds}. 
These workarounds often create a demand for new application designs and, in turn, new application requirements, in the case of SDP, for example, the trend towards higher-level language support.
However, at some point these emerging application designs will benefit more from new platforms that turn workarounds into supported platform features. Thus, new platform developments may be initiated, and new platforms emerge.

For the platform route in Figure~\ref{fig:method}, in a first step, we need to identify the application requirements that are better addressed through platform support. As described earlier, requirements such as the SDP requirements\cite{2018-Werner-BigData-MatrixMultiplication}, must be defined first. 
For later validation and to help with the identification of platform-driven limitations, experimental measurements and application~\cite{2021-Grambow-IC2E21-BeFaaS,2020-Werner-WoSC-SBDPs}- and platform benchmarks~\cite{2017-Bermbach-Book-CloudServiceBenchmarking,2020-Kuhlenkamp-ACR-All_But_One,Shahrad-2019-Micro-FaaSProfiler} to evaluate against these requirements are needed.
Based on the results, developers can either adapt their applications using the benchmarking results as a guide when designing necessary workarounds, or developers can start to implement new prototypical features in the platform and adapt the application to utilize these features accordingly.
By reusing or extending the application and platform benchmarks, we can evaluate if the changes lead to significant improvements for the application use case.
By iteratively applying these steps, we ultimately create a new platform adapted to the specific application needs or new applications that are adapted to current platform limitations.

\section{Conclusion}

In this paper, we described the application platform co-design phenomena and illustrated it for serverless computing platforms and serverless data processing, in particular.
More specifically, we discussed concrete challenges and needs in an SDP context and how these are initially addressed  through application workarounds, but may lead to new platform features and designs, resulting in a continuously changing platform landscape and the continuous need for developers to re-evaluate platforms and re-design applications.

The new SDP platform augmentations discussed have been implemented as part of the research project SMILE at TU Berlin for OpenWhisk~\cite{SMILE}.%
While we are still actively augmenting the platform to meet all the defined requirements for serverless data processing systems, we can already see significant improvements regarding function invocation management and processing throughput.

Through projects like SMILE and related work and observations, we expect more and more undesirable application workarounds to be eventually replaced by new platform features, confirming the continuous co-design phenomenon, but at the same time making clear, how little engineering support and understanding for such continuous co-design process exists to-date.
The duality of application and platform (re-)design challenges, the option to address identified limitations either on the application or the platform level, the continuous nature of both, and the need to study in depth fine-granular technical platform details, presents a larger challenge that demands new methods and tooling to better cope with application-platform co-design. We believe that application platform co-design awareness is critical to modern engineering needs such as SDP, and that appropriate methods and tooling should become an important piece of any developers tool-belt. Platforms will not stop evolving, and simultaneously, the choice of what software to use or adapt will grow, correspondingly.

\label{sec:conclu}

\bibliographystyle{splncs04}
\bibliography{references_noURL,icsoc}

\end{document}